\documentclass[aps,prc,twocolumn,superscriptaddress,nofootinbib,preprintnumbers,amsmath,amssymb,floatfix,showpacs]{revtex4-1}
\usepackage{graphicx}
\usepackage{amssymb}
\usepackage{dcolumn}
\usepackage{siunitx}
\usepackage{bm}
\usepackage{epsfig}
\usepackage{setspace}
\usepackage[caption=false]{subfig}
\usepackage{xcolor}
\begin{document}
\singlespacing

\preprint{}

\title{Dynamical formation of center domains in quark-gluon plasma}

\author{Jakapat Kannika}
\author{Christoph Herold}
\author{Ayut Limphirat}
\author{Chinorat Kobdaj}
\author{Yupeng Yan}

\affiliation{School of Physics, Suranaree University of Technology, Nakhon Ratchasima 30000, Thailand}

\date{\today}

\begin{abstract}

We study the formation of domain structures due to spontaneous breakdown of center symmetry at high temperatures in quenched QCD. We develop a phenomenological model for the explicit propagation of the Polyakov loop as the relevant order parameter of the deconfinement phase transition. The surface tension in the equation of motion is fit in comparison with lattice QCD data. Results give insight into the dynamical formation of center domains as well as the formation of energy bands along domain walls and let us estimate the required time to form such structures above the critical temperature.

\end{abstract}

\pacs{25.75.-q, 25.75.Nq, 11.30.Qc}

\maketitle

\section{Introduction}

The term quark-gluon plasma (QGP) denotes a state of matter, supposedly present in the early universe until about \SI{d-5}{\s} after the Big Bang. This primordial phase is characterized by the deconfinement of color charges and the restoration of chiral symmetry. Nowadays, the necessary energy densities to recreate a QGP are reached in heavy-ion collision experiments such as the Relativistic Heavy-Ion Collider (RHIC) at BNL or the Large Hadron Collider (LHC) at CERN. 

The transition from confinement to deconfinement is fairly well understood for pure gauge quantum chromodynamics (QCD), where it is possible to define an order parameter, the so-called Polyakov loop, related to the spontaneous breakdown of Z(3) center symmetry at high enough temperatures \cite{McLerran:1981pb}. For this case, lattice QCD data finds a first-order phase transition at $T_{\rm c}=\SI{270}{\MeV}$. With the inclusion of dynamical quarks, the transition temperature is significantly lowered to $T_{\rm c}=\SIrange[range-phrase=-]{150}{160}{\MeV}$ and the phase transition is smeared out to an analytic crossover \cite{Aoki:2006we,Borsanyi:2010bp}. At high temperatures, the Z(3) symmetric Polyakov loop potential has three degenerate minima, leading to a domain structure in the deconfined phase, where center symmetry spontaneously breaks into different gauge configurations in different spatial regions. 
Lattice QCD studies have confirmed the existence of these center domains for the SU(2) gauge group \cite{Fortunato:1999wr,Fortunato:2000ge,Fortunato:2002yn} and later for SU(3) \cite{Gattringer:2010ms,Danzer:2010ge,Borsanyi:2010cw,Endrodi:2014yaa}. The formation of such structures necessarily comes with domain walls, interpolating between the different values of Z(3) in neighboring domains \cite{Bhattacharya:1990hk,Bhattacharya:1992qb}.
Center domains have been claimed to provide a simultaneous explanation for two distinct properties of QGP, namely, the low ratio of shear viscosity over entropy density $\eta/s$ and jet quenching \cite{Asakawa:2012yv}, properties that have been experimentally confirmed at the LHC \cite{Muller:2012zq}.

Domain formation via bubble nucleation at the confinement-deconfinement transition has been studied within an effective model in \cite{Gupta:2010pp,Gupta:2011ag} both for quenched QCD and including dynamical quarks. In this work, the authors studied the evolution of QGP phase bubbles via the Kibble mechanism \cite {Kibble:1976sj,Zurek:1996sj} in a confining background in (2+1) dimensions, leading to Z(3) domains separated by domain walls and strings. It was shown that these structures cause inhomogeneities in the energy density even after cooling below $T_{\rm c}$, possibly influencing dilepton or direct photon distributions. It was argued furthermore that the expansion of these energetic fronts in the medium might leave  imprints in experimentally detectable flow coefficients. 

Our research here focuses on developing a dynamical model for an effective Polyakov loop field to describe the dynamical breaking of center symmetry and the formation of domains in the deconfined phase. For this purpose, we propose an effective Lagrangian consisting of a Polyakov loop potential from fits to lattice QCD and a phenomenological kinetic term, similar to what has been done in \cite{Dumitru:2000in,Fraga:2007gg,Herold:2013bi,Herold:2013qda}. We study the evolution of the Polyakov loop field in an isothermal heat bath in (3+1) dimensions, considering the relaxational dynamics after temperature quenches to the plasma phase. Effects of the heat bath are included by using a Langevin equation of motion. This enables us to include thermal fluctuations and thus compare results to recent lattice QCD studies at finite temperature \cite{Borsanyi:2010cw}. Comparison of the correlation length to the one obtained in \cite{Borsanyi:2010cw} allows us to fix the coefficient in front of the kinetic term in the region around $T_{\rm c}$. We are then able to study percolation of domains above the transition temperature and give estimates for formation times to better understand the possible role and relevance of these domain structure for heavy-ion collision experiments.

This paper is organized as follows: We begin with a description of the model in Sec.~\ref{sec:model}, followed by the numerical procedure in Sec.~\ref{sec:numerics}. In Sec.~\ref{sec:domain}, we fit the surface tension from lattice QCD correlation lengths and discuss formation procedures and estimate formation times in Sec.~\ref{sec:formation}. We conclude with a summary and outlook in Sec.~\ref{sec:summary}.

\section{Effective model}
\label{sec:model}

Confinement can be mathematically described by the Polyakov loop potential.
The fundamental Polyakov loop is defined as
\begin{equation}
\label{eq:polyakov_loop}
   L(\vec{x}) = \frac{1}{3}\mathrm{tr}P \exp\left[ig\int_{0}^{1/T} A_4(\tau, \vec{x}) d\tau\right]~,
\end{equation}

where $P$ denotes the path-ordering operator, $g$ is the strong-coupling constant, $T$ is the temperature, and $A_4$ is the temporal component of a static gluon background field in Euclidean space-time. 

From fits of lattice QCD data in the pure gluon sector, it is possible to obtain a polynomial potential for the Polyakov loop \cite{Wozar:2006fi,Heinzl:2005xv,Ratti:2005jh,deForcrand:2001nd,Pisarski:2000eq,Boyd:1996bx,Roessner:2006xn}. In our work, we use the version from \cite{Roessner:2006xn},

\begin{equation} \label{eq:potential_func}
U(L,T) = \left(-\frac{b_2}{2}\left| L\right|^2 -\frac{b_3}{6}( L^3+\bar L^3) +\frac{1}{4}(\left| L\right|^2)^2\right)b_4 T^4~,
\end{equation}

with the temperature-dependent coefficient 
\begin{equation} \label{eq:b2}
b_2(T)=((1-1.11/x)(1+0.265/x)^2(1+0.3/x)^3-0.487)/r^2~,
\end{equation}

and the parameters $b_3=2/r$, $b_4=0.61r^4$, where $x=T/T_c$ and $r=2.23$. This potential leads to an expectation value of $\langle L\rangle=0$ at temperatures $T<T_{\rm c}$ with $T_{\rm c}=\SI{270}{\MeV}$. Above $T_{\rm c}$, spontaneous symmetry breaking leads to three degenerate states $\langle L\rangle=\mathrm e^{i2\nu\pi/3}$ with $\nu=0,1,2$, the three elements of the center subgroup Z(3). The transition at $T_{\rm c}$ is of first-order type. In the presence of dynamical quarks, it becomes an analytic crossover and the Z(3) symmetry is also explicitly broken, preferring the state $\langle L\rangle=1$ at high temperatures. One can account for this by adding a term proportional to $L$ to the potential in Eq.~\eqref{eq:potential_func} as shown in \cite{Dumitru:2003cf,Meisinger:1995qr,Banks:1983me,Green:1983sd}. Note that as this potential is obtained from lattice QCD fits in equilibrium, it is well determined around the respective minima, but afflicted with uncertainties away from them. As discussed in \cite{Asakawa:2012yv}, the logarithmic form would lead to values of $L\approx 0$ along the domain walls representing gauge configurations similar to those in the confined phase. Within our dynamical model, we will demonstrate the emergence of these walls in Sec.~\ref{sec:formation}. 

Effective potentials for the Polyakov loop are often used in low-energy models such as the Polyakov loop Nambu-Jona-Lasinio (PNJL) model \cite{Fukushima:2003fw,Fukushima:2008wg} or the Polyakov-Quark-Meson (PQM) model \cite{Schaefer:2007pw,Herbst:2010rf}.

As the Polyakov loop carries no explicit time dependence, we apply an effective theory based on an effective Lagrangian of the form
\begin{equation}
\label{eq:free_energy}
 {\mathcal L}(L,T)=\frac{\sigma T_{\rm c}^2}{2}\left| \partial_\mu L \right|^2-U(L,T) ~,
\end{equation}
with the parameter $\sigma$ playing the role of a surface tension; cf.\ \cite{Fraga:2007gg}. It is clear that $\sigma$ will influence the domain size and therefore we shall assume it to be temperature-dependent, $\sigma\equiv\sigma(T)$. The factor of $T_{\rm c}^2$ has been added to account for the right dimensions in the fluctuation term \cite{Dumitru:2000in}. From the Lagrangian, we can obtain the equation of motion
\begin{equation}
\label{eq:eom}
 \sigma T_{\rm c}^2 \partial_\mu \partial^\mu L +\frac{\partial U(L,T)}{\partial L}=0 ~.
\end{equation}
To capture the effect of thermal fluctuations, we go beyond this classical equation by using a Langevin equation of motion including dissipation and noise in a thermalized heat bath:
\begin{equation}
\label{eq:eom2}
 \sigma T_{\rm c}^2 \partial_\mu \partial^\mu L +\eta \frac{\partial L}{\partial t}+\frac{\partial U(L,T)}{\partial L}=\xi~.
\end{equation} 
The dissipation coefficient $\eta$ has been estimated in \cite{Fraga:2007gg,Ananos:2007rj} by studying the exponential growth of the correlation function of the Polyakov loop within Glauber dynamics of pure SU(3) lattice gauge theory. Relating Monte Carlo time and real time as in \cite{Tomboulis:2005es} provides us with a value of $\eta=5/\textrm{fm}^3$. The stochastic noise field $\xi$ is Gaussian and white, and follows
\begin{equation}
\label{eq:dissfluct}
 \langle\xi(x,t)\xi(x',t')\rangle=2\eta T \delta(x-x')\delta(t-t')~.
\end{equation} 
from the dissipation-fluctuation theorem. 
From Eqs.~\eqref{eq:free_energy} and \eqref{eq:eom}, it is clear that the correlation length $\xi$ depends on the coefficient $\sigma$ as $\xi\sim\sqrt{\sigma}$. It has been shown in \cite{Gattringer:2010ms,Borsanyi:2010cw} that in quenched QCD, the cluster size $d=2\xi$ is constant below $T_{\rm c}$ at about \SI{0.5}{\femto\metre} and then linearly rising above $T_{\rm c}$, thus signaling the phase transition. The authors argued that the value of \SI{0.5}{\femto\metre} proves reasonable as it resembles the size of a heavy quark meson. We therefore assume a quadratic increase of $\sigma$ with temperature above the phase transition and use the following ansatz for the kinetic coefficient:
\begin{equation}
\label{eq:sigmaT}
 \sigma=
 \begin{cases} \sigma_0 &\mbox{if } T\leq T_{\rm c} \\
\sigma_0+a\left(\frac{T-T_{\rm c}}{T_{\rm c}}\right)^2 & \mbox{if } T>T_{\rm c} \end{cases}~.
\end{equation}
In a next step, we fix the constants $\sigma_0$ and $a$ from the simulation by extracting the correlation length $\xi$ and comparing it with the values obtained in \cite{Borsanyi:2010cw}. The ansatz in Eq.~\eqref{eq:sigmaT} is of course an oversimplification as we neglect the curvature of the Polyakov loop potential which also influences the correlation length. It is also not clear if the linear rise in the cluster size is at some point halted as the data provided in \cite{Borsanyi:2010cw} only reaches up to $1.2\,T_{\rm c}$.

\section{Numerical setup}
\label{sec:numerics}

To study the evolution of the center domains, we numerically solve the equation of motion \eqref{eq:eom2} in (3+1) dimensions. We apply periodic boundary conditions for the spatial coordinates on a cubic lattice with $200^3$ sites, large enough to contain several larger domains at higher temperatures. 
We correlate the noise field over a volume of $(1/T)^3$ for each box with given temperature $T$ as we expect the correlation length to be of the order of $1/T$. We chose the lattice spacing to be $\Delta x N_{\rm corr}=1/T$ with $N_{\rm corr}=4$, following \cite{Bazavov:2008qh} such that for each box, the noise field is correlated over a volume of $N_{\rm corr}^3$ cells. Finally, we set the time step to $\Delta t=0.01\cdot\Delta x$ to ensure numerical stability. 
At time $t=\SI{0}{\femto\metre}$, the Polyakov loop $L=|L|\mathrm e^{i\phi}$ is initialized with a flat distribution in its argument $\phi$ and a Gaussian with width $0.1$ in the modulus $|L|$, corresponding to small fluctuations around the low-temperature expectation value of $L=0$. Note that Eq.~\eqref{eq:eom} does not include the Hubble term; thus we neglect the effects of an expanding medium and consider a static box only.

\section{Domain size}
\label{sec:domain}

\begin{figure}[t]
\centering
    \includegraphics[scale=1.2]{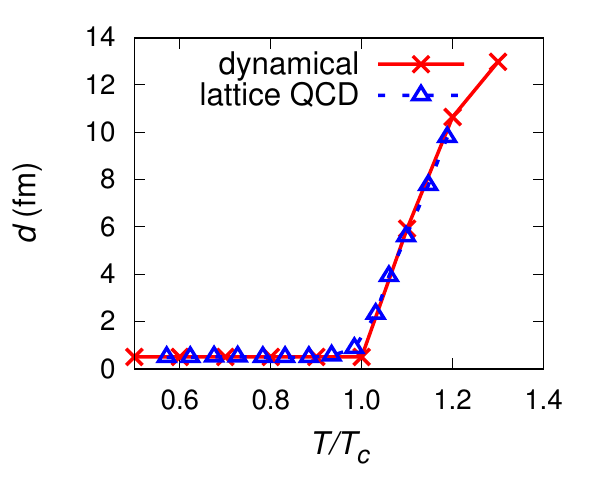}
\caption[dphys]{(Color online) Correlation length of the Polyakov loop from our dynamical model compared to the results obtained on the lattice in \cite{Borsanyi:2010cw}.}
\label{fig:dphys}
\end{figure}

Our first goal is to determine the behavior of the kinetic coefficient $\sigma$ as a function of $T$. We use the temperature values studied in \cite{Borsanyi:2010cw} as quench temperatures above $T_{\rm c}$ and then determine the correlation length $\xi$ from the two-point correlation function of the imaginary part $L_2$ of the Polyakov loop $L=L_1+iL_2$,
\begin{equation}
\label{eq:corr}
C(\left|x-y\right|)=\langle L_2(x)L_2(y)\rangle_T\propto \mathrm e^{-\left|x-y\right|/\xi}~,
\end{equation}
taking the average $\langle\dots\rangle_T$ over an ensemble with the same temperature $T$. The value of $L_2$ serves to distinguish different types of domains which are characterized by $L_2=0$, $L_2=\sin(2\pi/3)$, and $L_2=-\sin(2\pi/3)$. For the real part $L_1$, only two different values occur. Alternatively, one could use the argument $\phi$ of $L$. 

After setting the temperature in our box to the desired value above $T_{\rm c}$, we follow the evolution of the system to the deconfined phase, where the small initial fluctuations amplify and form center domains. As soon as the evolution has come to a halt and no more domains form or merge, we determine $\xi$ from the correlation function \eqref{eq:corr}. We tune our input $\sigma$ such that the cluster diameter $d=2\xi$ equals the equally obtained $d$ from the lattice calculation \cite{Borsanyi:2010cw}. From a fit of $\sigma$ as function of $T$ according to Eq.~\eqref{eq:sigmaT}, we finally obtain the values $\sigma_0=0.004$ and $a=4.3$. 
Next, we use $\sigma(T)$ to calculate the domain size for several values of $T$, ranging from $0.5\,T_{\rm c}$ to $1.3\,T_{\rm c}$, and show the result in Fig.~\ref{fig:dphys}. We see a clear resemblance between the data from our dynamical model and the lattice QCD results in the limited range investigated in \cite{Borsanyi:2010cw}. In both cases, the phase transition is clearly indicated by the sudden increase in $d$. For temperatures above the provided lattice QCD data, the cluster diameter from our model no longer increases linearly with temperature. This is due to the obvious oversimplification in our ansatz~\eqref{eq:sigmaT}, where we neglected the curvature of the potential as an influence on the correlation length. Calculations of the volume of the largest percolating cluster in \cite{Endrodi:2014yaa} show that above $T_{\rm c}$, this volume increases less strongly with growing temperature. We may therefore expect a similar behavior for the average diameter, thus at least qualitatively justifying our result.

\section{Dynamical domain formation}
\label{sec:formation}

\begin{figure}[]
\centering
   \includegraphics[scale=0.7]{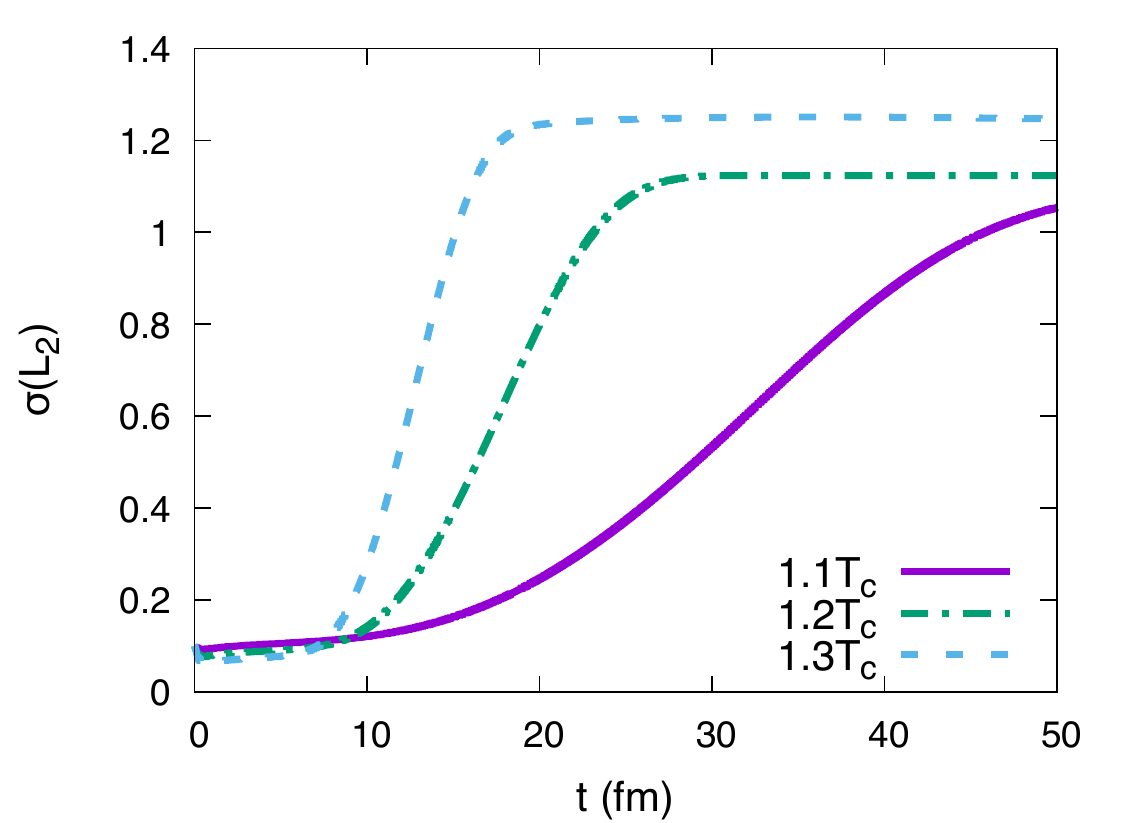}
\caption[sigma]{(Color online) Time evolution of the standard deviation of $L_2$, indicating the formation of center domains in different quench scenarios.}
\label{fig:sigma}
\end{figure}

To give an estimate for the formation time of domains, we use the standard deviation of $L_2$ over all cells in our box, 
\begin{equation}
\sigma(L_2)=\sqrt{\langle(\delta L_2)^2\rangle_V}~,
\end{equation}
with $\delta L_2=L_2-\langle L_2\rangle_V$ and the average $\langle\dots\rangle_V$ taken over all cells in our volume $V$. We expect $\sigma(L_2)$ to saturate at some value as soon as the dynamical formation of domains in the isothermal heat bath has settled. We show this quantity as a function of time in Fig.~\ref{fig:sigma} for several quench temperatures. Note here that the temperature is increased directly at $t=\SI{0}{\femto\metre}$. We obtain values in the range between $17.0$ and $\SI{50.0}{\femto\metre}$ for the formation. This formation time increases with temperature as with increasing surface tension $\sigma$, the velocity $\partial L/\partial t$ decreases. We point out that the obtained curves lie closer and closer with increasing quench temperature, indicating a slowing down in the increase of the formation time. For center domains to play a relevant role in heavy-ion experiments, one needs to consider two things: First, the average domain volume should be smaller than the volume of the medium created after the collision, and second, the formation time has to be smaller than the lifetime of the QGP. Both quantities have been determined by Hanbury-Brown-Twiss (HBT) measurements \cite{Aamodt:2011mr} who found a homogeneity volume at LHC of around $\SI{300}{\femto\metre}^3$, twice the size as at RHIC. This volume would be enough to contain several smaller domains, given the average domain size does not exceed about \SI{7}{\femto\metre} for LHC or \SI{5.5}{\femto\metre} for RHIC. The decoupling time has been determined to reach up to \SIrange[range-phrase=--]{10}{11}{\femto\metre} at LHC energies in comparison to \SIrange[range-phrase=--]{7}{8}{\femto\metre} at RHIC, well in agreement with previous transport model simulations \cite{Bass:2000ib}. Therefore, taking into account our estimates for the required formation time, the possibility of domain formation at the LHC seems doubtful. However, it would be favorable to extend results to higher temperatures and also consider an expanding medium to simulate the situation in a heavy-ion collision. Work in this direction is currently in progress. 

\begin{figure*}[t]
\centering
    \subfloat[\label{fig:t4}]{
    \centering
    \includegraphics[scale=0.9]{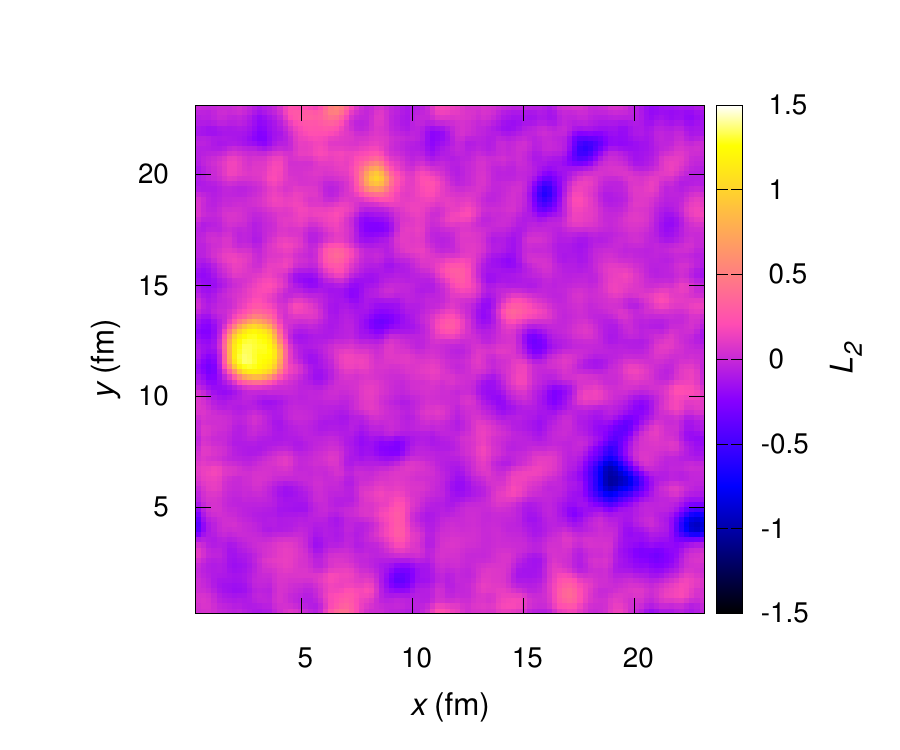}
    }
  \hfill
    \subfloat[\label{fig:t6}]{
    \centering
    \includegraphics[scale=0.9]{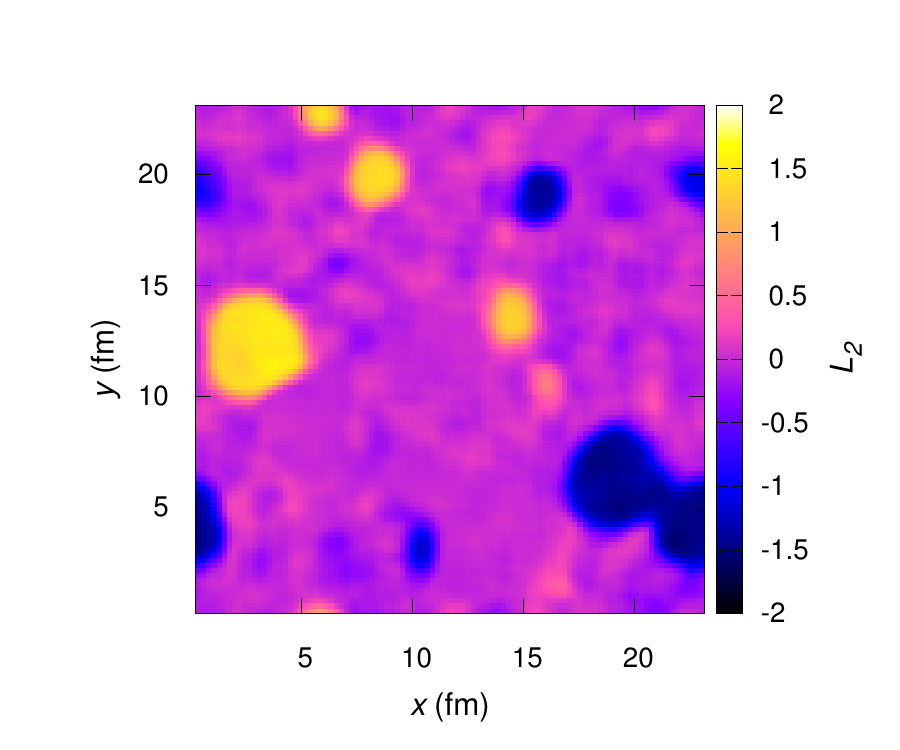}
    }
    \vfill
   \subfloat[\label{fig:t8}]{
    \centering
    \includegraphics[scale=0.9]{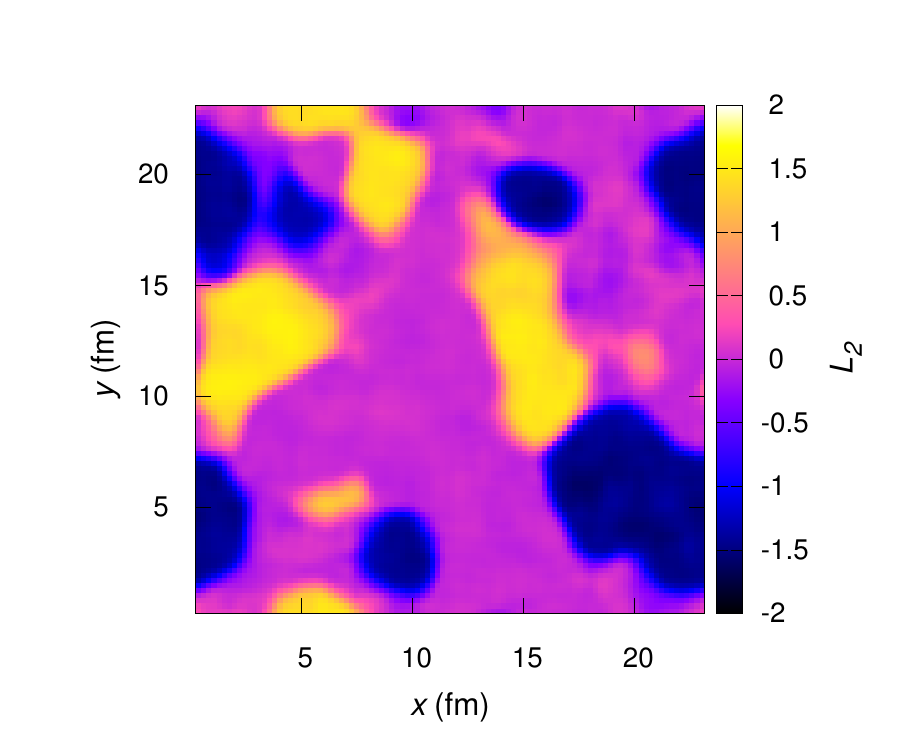}
    }
  \hfill
    \subfloat[\label{fig:t12}]{
    \centering
    \includegraphics[scale=0.9]{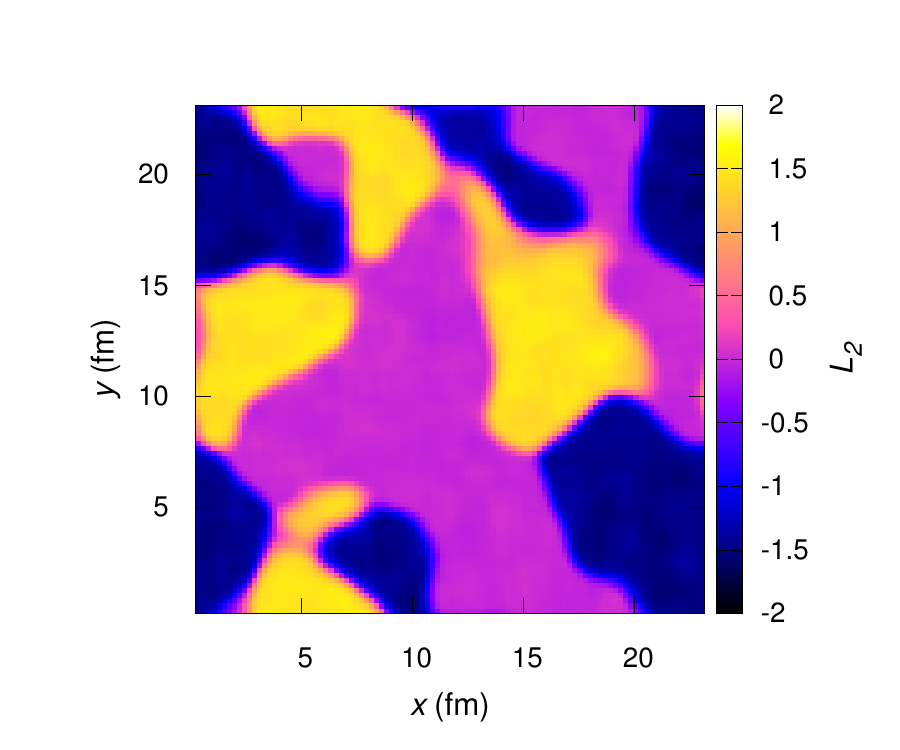}
    }
\caption[maps]{(Color online) The imaginary part of the Polyakov loop for constant $z$ at $T=1.1\,T_{\rm c}$ and times $t=\SI{20}{\femto\metre}$ \subref{fig:t4}, $t=\SI{30}{\femto\metre}$ \subref{fig:t6}, $t=\SI{40}{\femto\metre}$ \subref{fig:t8}, $t=\SI{50}{\femto\metre}$ \subref{fig:t12}. The formation and merging of domains can be observed clearly.}
\label{fig:maps}
\end{figure*}

\begin{figure}[t]
\centering
    \includegraphics[scale=0.9]{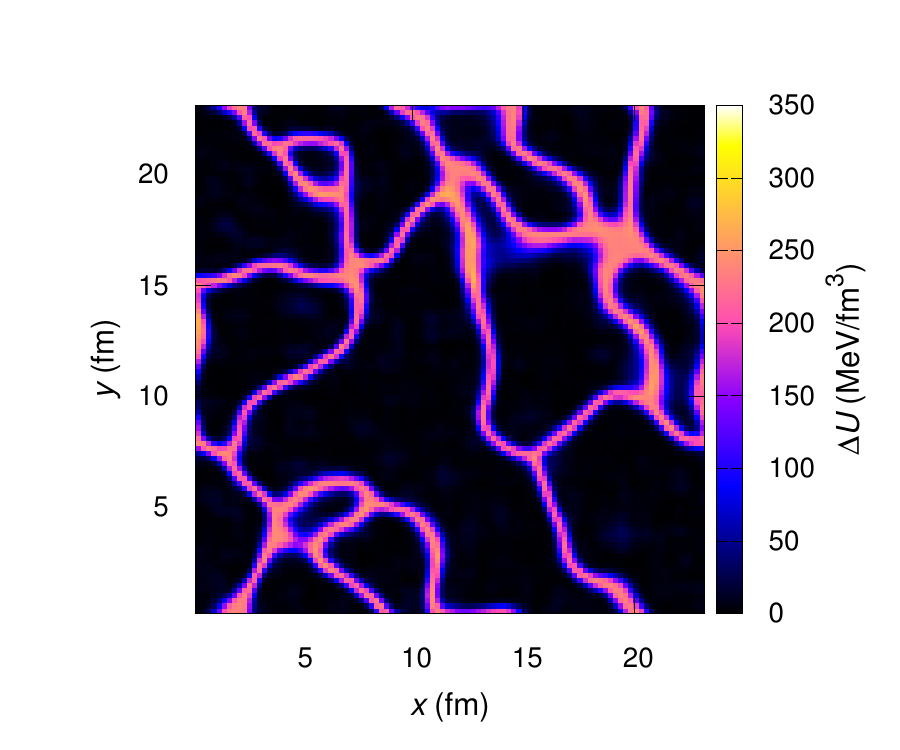}
\caption[border]{(Color online) Energy borders on the surfaces of the center domains at $T=1.1\,T_{\rm c}$ and $t=\SI{50}{\femto\metre}$.}
\label{fig:border}
\end{figure}

It is instructive to follow a visualization of the evolution of center domains. We show plots of the imaginary part of $L$ in Fig.~\ref{fig:maps} for constant value of $z$, corresponding to the curve of $T=1.1\,T_{\rm c}$ in Fig.~\ref{fig:sigma}. Center domains then appear as regions of the same color. We choose four distinct times, namely, at the beginning of the steep increase of $\sigma(L_2)$ ($t=\SI{20}{\femto\metre}$), during the increase ($t=\SI{30}{\femto\metre}$), at its end ($t=\SI{40}{\femto\metre}$), and at the end of the simulation ($t=\SI{50}{\femto\metre}$). In the first two plots, the shape of the domains becomes more distinct while the field values move towards their respective equilibrium values. In the next two plots, we see the domains further sharpen in the edges and observe some domains merging with neighboring domains of the same type. It is important to note that these domains are all continuously connected and there is no abrupt change in the values, but a smooth transition between volumes with different Z(3) configuration. This border is characterized by a gauge configuration corresponding to the confined low-temperature phase, as already discussed in Sec.~\ref{sec:model}. We demonstrate that this different gauge configuration effectively leads to a difference in the energy (see Fig.~\ref{fig:border}), which shows the energy density $\Delta U$ above the ground state at time $t=\SI{50}{\femto\metre}$, corresponding to the plot of $L_2$ in Fig.~\ref{fig:t12}. Here, we clearly see bands of high energy at the edges of the center domains.

This is an important observation as these borders might act as potential barriers in two ways: Soft partons with thermal momenta will reflect on the domain walls which effectively limits their free wave length, resulting in a small value of $\eta/s$. Hard partons, on the other hand, may cross these walls under the emission of soft gluon radiation. Reflection of these gluons on the walls then makes the jet energy rapidly isotropic \cite{Asakawa:2012yv}. It would be interesting in the future to extend our model such that it can describe these two effects and estimate their impact on shear viscosity and jet quenching.

\section{Summary and Outlook}
\label{sec:summary}

In this article, we have introduced an effective model to describe the formation of Z(3) center domains via dynamical symmetry breaking in pure gluon QCD. We have fixed the surface tension in our equation of motion to obtain correlation lengths corresponding to the latest lattice QCD results. Our model signals the phase transition via a sudden increase in the size of center domains. We observed the mechanism of domain formation after a temperature quench, through the formation of small bubbles and the subsequent merging of bubbles with the same gauge configuration into domain structures. We were able to give estimates for the formation time, ranging from \SIrange[range-phrase=--]{17}{50}{\femto\metre}, larger than the estimated lifetime of the QGP at LHC energies. Although the presented results cover only a limited range up to $1.3\,T_{\rm c}$, we see that the formation time increases less and less with increasing temperature. 

Finally, we have demonstrated the occurrence of energy bands along the walls of domains, which have previously been claimed to be a possible explanation for the simultaneous emergence of low $\eta/s$ and jet quenching. 

We are going to continue our work by considering the effects of dynamical quarks though a linear term in the effective potential for the Polyakov loop, indicating the explicit symmetry breaking present in full QCD. We expect that in this case, domain formation will also occur, as lattice QCD calculations with dynamical quarks also lead to the formation of similar structures above the transition temperature \cite{Deka:2010bc}. Furthermore, dynamical models have also found domain formation if explicit symmetry breaking is taken into account \cite{Mohapatra:2012ck}. Without a first-order phase transition, it can be expected that near $T_c$, the transition between the two phases proceeds more rapidly in a dynamical setup as there is no barrier to overcome, which significantly decreases the formation time for these scenarios. 
We are going to continue our work by considering the effects of dynamical quarks though a linear term in the effective potential for the Polyakov loop. To study the situation after the collision of two nuclei, it would furthermore be instructive to couple this model to a full (3+1)-dimensional hydrodynamical expansion to better understand what happens to center domains during expansion and cooling.

\section*{Acknowledgements}

This work is funded by Suranaree University of Technology (SUT) and by the Office of the Higher Education Commission under NRU project of Thailand. C.H. and C.K. acknowledge support from SUT-CHE-NRU (FtR.15/2559) project.

The computing resources have been provided by the National e-Science Infrastructure Consortium of Thailand,
the Center for Computer Services at SUT, and the Frankfurt Center for Scientific Computing.

\section*{References}
\bibliography{mybib}

\end{document}